# Classical and Quantum Mechanical (QM) Effects in the One-Soliton Solution of the EM Nonlinear Schrodinger (NLS) Equation

Y.Ben-Aryeh

Physics Department, Technion Israel of Technology, Haifa 32000, Israel

E-mail address: phr65yb@physics.technion,ac.il;  Fax:972 4 8295755

**Abstract**

Propagation effects are analyzed for electromagnetic (EM) waves which satisfy the one-soliton non-linear Schrodinger (NLS) equation in a dispersive wave guide. The coupling between momentum and frequencies due to dispersion relation is treated by *a coupled Hamiltonian-Momentum operator* with equal-space commutation relations (CR). Kerr interactions in the soliton are treated. Photon number distribution and quantum phases are analyzed. The integrability of nonlinear equations can be related to *compatibility-condition* between certain Hamiltonian and Momentum  matrices and such relation is applied to the NLS equation.

# 1. Introduction

In the present study we would like to analyze classical and quantum mechanical (QM) effects for electromagnetic (EM) waves which satisfy the one-soliton non-linear Schrodinger (NLS) equation in a dispersive wave guide using quantum optics methods. There is an extensive literature on various solutions for NLS equation including especially soliton solutions (e.g. [1-11]). The common procedure for studying QM effects for the NLS equation is by the use of quantum field theory [4-5,9]. In such approach one uses the field operators $\hat{\phi}(z,t)$ and $\hat{\phi}^{\dagger}(z,t)$ satisfying the boson commutation relations (CR)

$$\left[\hat{\phi}(z,t), \hat{\phi}^{\dagger}(z',t)\right] = \delta(z-z') \quad , \tag{1}$$

where $z$ is the one dimensional propagation coordinate. For one dimensional Hamiltonians one uses multiplications of such operators and their derivatives integrated over the $z$ coordinate. Usually QM solutions of NLS equation by using quantum field theories turn to be quite complicated (see e.g. Ref. [9]). In Quantum Optics one can use the conventional one dimensional equal time CR for the annihilation and creation operators given as

$$\left[\hat{a}(k,t), \hat{a}^{\dagger}(k',t)\right] = \delta(k-k') \tag{2}$$

where $k$ and $k'$ are the one dimensional wave vectors. These CR can be converted into another form by using the following Fourier transforms

$$\hat{a}(z,t) = \frac{1}{\sqrt{2\pi}} \int_{-\infty}^{\infty} \exp(ikz)\hat{a}(k,t)dk \quad ; \quad \hat{a}^{\dagger}(z',t) = \frac{1}{\sqrt{2\pi}} \int_{-\infty}^{\infty} \exp(-ik'z')\hat{a}^{\dagger}(k',t)dk' \quad .(3)$$

Then we get

$$\left[\hat{a}(z,t), \hat{a}^{\dagger}(z',t)\right] = \frac{1}{2\pi}\int_{-\infty}^{\infty}\int_{-\infty}^{\infty}\delta(k-k')\exp(ikz-ik'z')dkdk' =$$

$$\frac{1}{2\pi}\int_{-\infty}^{\infty}\exp\{ik(z-z')\}dk = \delta(z-z') \quad . \tag{4}$$

This CR will be used in the present analysis of QM effects related to the NLS equation.

There are other forms for CR in quantum optics (see e.g. [12]). For various systems the momentum operator $\hat{G}$ has been used for evaluating quantum mechanical (QM) propagation effects including coupling between different momentum modes (see e.g. [13-17]). In the present work we treat, however, one mode of the EM field which includes a coupling between momentum and frequencies due to dispersion relation. For this purpose it is useful to treat propagation effects by *a coupled Hamiltonian-Momentum operator* and to use the CR of Eq. (4) as it is the analog of the quantum field CR of Eq. (1).

Quantum effects obtained by Kerr interaction have been treated in various articles [16-20]. These works have not treated, however, the Kerr propagation effects in solitons. Haus [8] and Haus and Lai [21] have developed the quantum theory of soliton squeezing using a linearized approach. Although some of the derivations in the present work are similar to those presented in [8] the approach for treating Kerr effects is different and more similar to other conventional quantum optics methods [22], especially for treating quantum phases in the soliton. Although we will treat QM effects for the special important case of *one-soliton solution* of NLS equation, a similar QM approach can be used for more complicated solutions of NLS equation.

The article is arranged as follows. In Section 2, QM propagation effects for *linear* dispersive wave guide are treated. In Section 3 we analyze the one-soliton solution of the NLS equation in which nonlinear Kerr interaction effects are taken into account. In conventional QM analysis $\hat{a}^\dagger(t)\hat{a}(t)$ represents the photon number operator in the quantization volume while in our analysis $\hat{a}^\dagger(z,t)\hat{a}(z,t)$ represents the number of photons per unit length at coordinates $z,t$. The total photon number operator $\hat{n}$ in the soliton is given by $\int \hat{a}^\dagger(z,t)\hat{a}(z,t)dz = \hat{n}$. Based on these properties we analyze in Section 3 possible QM phases in the one soliton solution of the NLS equation and compare our results with other works. The main point emphasized by the analysis given in this Section is that soliton solutions can be related to QM annihilation $\hat{a}(z,t)$ and creation $\hat{a}^{\dagger}(z,t)$ operators from which quantum phases can be calculated. This approach for the one-soliton case is different from that presented in [8,21] and is more directly related to the ordinary treatment of quantum phenomena by Kerr interactions in other systems [14-18]. In the present analysis we do not consider more complicated phenomena, e.g. losses, diffraction at the transverse coordinate, multi-soliton solutions, collisions between solitons etc. [1-2,10-11].

Nonlinear equations raise the problem how far they are integrable. We show in Section 4 that the NLS equation can be related to two first order linear equations operating on a two-dimensional wavefunction [1-2,6-7,23], where its time dependence is given by 'Hamiltonian -Matrix' while its space $z$ dependence is given by 'Momentum-Matrix'. The integrability of the nonlinear equations is related to the *compatibility-condition* between these two matrices and the use of such *compatibility-condition* is demonstrated for the NLS equation. We do not develop in the present article the use of scattering theories as they are usually classical theories (e.g.[1-2], [6-7]) and our aim in the present article is to treat QM Kerr effects which can be done by the methods developed in the Section 3.

In Section 5 we summarize our results and conclusions.

## 2. Propagation of EM waves in a linear dispersive wave guide

In treating EM waves in *linear dispersive* wave guide we encounter the problem that the frequency $\omega$ and the wavevector k are coupled by the dispersion relation [8]:

$$\omega(k) \simeq \omega_0(k_0) + \left(\frac{d\omega}{dk}\right)_{k=k_o} \delta k + \left(\frac{d^2\omega}{dk^2}\right)_{k=k_o} \delta k^2 \quad . \tag{5}$$

Here $\omega_0$ is the resonant frequency corresponding to the wavevector $k_0$,

$$\left(\frac{d\omega}{dk}\right)_{k=k_o} = v_g \tag{6}$$

is the group velocity, and $\left(\frac{d^2\omega}{dk^2}\right)_{k=k_o}$ is the group velocity dispersion. Higher order terms in the series expansion of $\omega$ as function of $\delta k$ are neglected here as we assume that we have only a narrow distribution of wavevectors around the central wavevector $k_0$. We limit the analysis given in the present Section to the one dimensional case.

The coupled *Hamiltonian-Momentum operator* which includes *the coupling between frequencies and wavevectors* is given as

$$\hat{H} = \hbar \int \omega(k_0 + \delta k)\hat{a}^\dagger(k_0 + \delta k)\hat{a}(k_0 + \delta k)d(\delta k) \tag{7}$$

By substituting Eq. (5) into Eq. (7) we get:

$$\hat{H} = \hbar \int \left[\omega_0(k_0) + \left(\frac{d\omega}{dk}\right)_{k=k_o}\delta k + \left(\frac{d^2\omega}{dk^2}\right)_{k=k_o}\delta k^2\right]\hat{a}^\dagger(k_0 + \delta k)\hat{a}(k_0 + \delta k)d(\delta k) . \tag{8}$$

In Eq. (8) $\hat{a}^\dagger(k_0 + \delta k)\,\hat{a}(k_0 + \delta k)$ represents the number of photons which have momentum $k_0 + \delta k$ with the energy given in the square brackets, and this energy is summed over all modes represented by the integration over $\delta k$. We will transform the dependence of the creation and annihilation operators in Eq. (8) on the momentum $k_0 + \delta k$ to dependence on the coordinate $z$ by using the following Fourier transforms:

$$\hat{a}(k_0 + \delta k) = \frac{1}{\sqrt{2\pi}}\int \hat{a}(z)\exp\left[i(k_0 + \delta k)z\right]dz \;;$$

$$\hat{a}^\dagger(k_0 + \delta k) = \frac{1}{\sqrt{2\pi}}\int \hat{a}^\dagger(z')\exp\left[-i(k_0 + \delta k)z'\right]dz' \;; \tag{9}$$

$$\hat{a}^\dagger(k_0 + \delta k)\hat{a}(k_0 + \delta k) = \frac{1}{2\pi}\int dz \int dz'\, \hat{a}^\dagger(z')\hat{a}(z)\exp\left[i(k_0 + \delta k)(z - z')\right]$$

where the operators $\hat{a}(z)$ and $\hat{a}^\dagger(z)$ satisfy the CR given in Eq. (4) which in short notation can be written as:

$$\left[\hat{a}(z), \hat{a}^\dagger(z')\right] = \delta(z-z') \quad . \tag{10}$$

We substitute Eq. (9) into Eq. (8) obtaining

$$\hat{H}/\hbar =$$
$$\int d(\delta k) \left[\omega_0 + \left(\frac{d\omega}{dk}\right)_{k=k_o} \delta k + \left(\frac{d^2\omega}{dk^2}\right)_{k=k_o} \delta k^2\right] \iint dz dz' \hat{a}^\dagger(z')\hat{a}(z) \exp\left[i(k_0+\delta k)(z-z')\right]$$

$$\tag{11}$$

and perform the following three integrals:

$$\frac{1}{2\pi}\int \exp\left[i(k_0+\delta k)(z-z')\right] d(\delta k) = \delta(z-z') \quad , \tag{12}$$

$$\frac{1}{2\pi}\int\left\{\left(\frac{d\omega}{dk}\right)_{k=k_o}\delta k\right\}\exp\left[i(\delta k)(z-z')\right]d(\delta k) =$$
$$-\frac{i}{2\pi}\left(\frac{d\omega}{dk}\right)_{k=k_o}\frac{\partial}{\partial z}\int \exp\left[i(\delta k)(z-z')\right]d(\delta k) = -i\left(\frac{d\omega}{dk}\right)_{k=k_o}\frac{\partial}{\partial z}\delta(z-z')$$
$$, \tag{13}$$

$$\frac{1}{2\pi}\int\left\{\left(\frac{d^2\omega}{dk^2}\right)_{k=k_o}\delta k^2\right\}\exp\left[i(\delta k)(z-z')\right]d(\delta k) =$$
$$-\frac{1}{2\pi}\left(\frac{d^2\omega}{dk^2}\right)_{k=k_o}\frac{\partial^2}{\partial z^2}\int \exp\left[i(\delta k)(z-z')\right]d(\delta k) = -\left(\frac{d^2\omega}{dk^2}\right)_{k=k_o}\frac{\partial^2}{\partial z^2}\delta(z-z')$$
$$.\tag{14}$$

Substituting Eqs. (12-14) into Eq. (11) and transforming the derivatives by partial integration we get:

$$\hat{H} =$$
$$\hbar\omega_0\int \hat{a}^\dagger(z')\hat{a}(z')dz' - i\hbar\left(\frac{d\omega}{dk}\right)_{k=k_o}\int \hat{a}^\dagger(z')\left\{\frac{\partial}{\partial z'}\hat{a}(z')\right\}dz' - \hbar\left(\frac{d^2\omega}{dk^2}\right)_{k=k_o}\int \hat{a}^\dagger(z')\left\{\frac{\partial^2}{\partial z'^2}\hat{a}(z')\right\}dz'$$

$$\tag{15}$$

In Eq. (15) we arranged the space derivative so that they operate to the right. By using partial integration Eq. (15) can be transformed into the other form

$$\hat{H} = \hbar\omega_0 \int \hat{a}^\dagger(z')\hat{a}(z')dz' + i\hbar \left(\frac{d\omega}{dk}\right)_{k=k_o} \int \left\{\frac{\partial}{\partial z'}\hat{a}^\dagger(z')\right\}\hat{a}(z')dz' - \hbar\left(\frac{d^2\omega}{dk^2}\right)_{k=k_o} \int \left\{\frac{\partial^2}{\partial z'^2}\hat{a}^\dagger(z')\right\}\hat{a}(z')dz'$$

. (16)

The *Hamiltonian* of Eq. (15) or of Eq. (16) is the generator for the coupled space-time propagation. The equation of motion for the annihilation operator is given by

$$i\hbar \frac{\partial}{\partial t}\hat{a}(z,t) = \left[\hat{a}(z,t), \hat{H}\right] \quad , \tag{17}$$

where $\hat{H}$ is given by Eq. (15) and the CR of Eq. (4) can be used. By substituting Eq. (15) into Eq. (17) and performing the CR we get

$$\frac{\partial}{\partial t}\hat{a}(z,t) = -i\left(\omega_0 \hat{a}(z,t) - i\left(\frac{d\omega}{dk}\right)_{k=k_o}\frac{\partial}{\partial z}\hat{a}(z,t) - \left(\frac{d^2\omega}{dk^2}\right)_{k=k_o}\frac{\partial^2}{\partial z^2}\hat{a}(z,t)\right) . \tag{18}$$

The corresponding equation for $\frac{\partial}{\partial t}\hat{a}^\dagger(z)$ can be obtained by the dagger of Eq. (18) (or correspondingly from Eq. (16)).

We define:

$$\hat{a}(z) = \tilde{\hat{a}}(z)\exp(-i\omega_0 t) \quad , \tag{19}$$

and then Eq. (18) is transformed into

$$\frac{\partial}{\partial t}\tilde{\hat{a}}(z,t) = -v_g \frac{\partial}{\partial z}\tilde{\hat{a}}(z,t) + i\frac{1}{2}\left(\frac{d^2\omega}{dk^2}\right)_{k=k_o}\frac{\partial^2}{\partial z^2}\tilde{\hat{a}}(z,t) \quad , \tag{20}$$

where $v_g$ is the group velocity defined in Eq. (6).

We use the following additional transformation:

$$t' = t - \frac{z}{v_g} \quad , \tag{21}$$

and then Eq. (20) is transformed into

$$\frac{\partial}{\partial t'}\tilde{\hat{a}}(z,t) = i\frac{1}{2}\left(\frac{d^2\omega}{dk^2}\right)_{k=k_o}\frac{\partial^2}{\partial z^2}\tilde{\hat{a}}(z,t) \qquad . \tag{22}$$

The time $t' = t - \frac{z}{v_g}$ indicates that we removed the time delay $\frac{z}{v_g}$ from the ordinary time and the use of operator $\tilde{\hat{a}}(z)$ indicates that we removed the rapid carrier oscillation frequency $\exp(-i\omega_0 t)$ from $\hat{a}(z)$.

One should notice that the Hamiltonian of Eq. (15) or (16) includes integration over the $z'$ coordinate where such integration is in analogy with quantum field Hamiltonians which include also such space integration (see e.g [4-5, 9]). Our derivations are, therefore, based on the use of the CR of Eq. (4).

For simplicity of notation, from now on, in using Eq. (22) we will remove the prime and the tilde from this equation but we need to take into account that the time in such equation represents a time delayed by $\frac{z}{v_g}$ and that from $\hat{a}(z)$ we removed the rapid variation $\exp(-i\omega_0 t)$.

## 3. The Classical and QM one-soliton solution of the NLS equation

The nonlinear Schrodinger Hamiltonian can be obtained by adding to the linear *Hamiltonian* of Eq. (11) or that of Eqs. (15-16) the Kerr effect represented by *nonlinear Hamiltonian*

$$\hat{H}_K(t) = -\hbar\frac{K}{2}\int dz\hat{a}^\dagger(z,t)\hat{a}^\dagger(z,t)\hat{a}(z,t)\hat{a}(z,t) = -\hbar\frac{K}{2}\int dz\hat{n}(z,t)\left(\hat{n}(z,t)-1\right), \tag{23}$$

where K is the Kerr constant and $\hat{n}(z,t) = \hat{a}^{\dagger}(z,t)\hat{a}(z,t)$) is the number of photons per unit length at coordinates $(z,t)$.

By taking into account the nonlinear momentum operator $\hat{H}_K$ the equation of motion for $\hat{a}(z,t)$ becomes

$$\frac{\partial}{\partial t}\hat{a}(z,t) = i\frac{1}{2}\left(\frac{d^2\omega}{dk^2}\right)_{k=k_o}\frac{\partial^2}{\partial z^2}\hat{a}(z,t) - \frac{i}{\hbar}\left[\hat{a}(z), \hat{H}_k(z,t)\right] = i\frac{1}{2}C\frac{\partial^2}{\partial z^2}\hat{a}(z,t) + iK\hat{a}^{\dagger}(z,t)\hat{a}(z,t)\hat{a}(z,t)$$

(24)

We have used here Eq. (22) where for simplicity of notation we have omitted the tilde and prime from this equation (see the above explanation), and added the CR with $\hat{H}_K$. Also we have defined in a short notation the constant $C = \left(\frac{d^2\omega}{dk^2}\right)_{k=k_o}$ representing the group velocity dispersion.

In the classical description we exchange the operator $\hat{a}(z,t)$ into its classical representation $a_c(z,t)$ and then we get the NLS classical equation

$$\frac{\partial}{\partial t}a_c(z,t) = i\frac{1}{2}C\frac{\partial^2}{\partial z^2}a_c(z,t) + iK\hat{a}_c^{*}(z,t)a_c(z,t)a_c(z,t) \qquad . \qquad (25)$$

The one-soliton solution of the NLS equation depends on 4 constants: The constant $x_0$ representing the pulse center, the carrier frequency of the soliton, arbitrary phase constant $\theta_0$ of the soliton and the total intensity of the soliton. The solution of Eq. (24) is simplified by eliminating these constants [7,8]: The constant $x_0$ representing the soliton center is eliminated by choosing a coordinate system whose origin is at the pulse center. The arbitrary constant phase $\theta_0$ of $a_c(z,t)$ is chosen to be equal to zero. The carrier frequency of the soliton is assumed to coincide with the carrier frequency $\omega_0$. The total number of photons in the soliton remains as an important parameter. Under these assumptions the soliton classical solution of Eq. (25) is given as

$$a_c(z,t) = A \exp\left[i\left(\frac{KA^2}{2}t\right)\right] \sec h\left(\frac{z}{\xi}\right) \quad , \tag{26}$$

with the constraint

$$KA^2 = \frac{C}{\xi^2} \quad . \tag{27}$$

The normalization constant $A$ is assumed to be real. The complex classical amplitude $a_c(z,t)$ is normalized so that by integrating its magnitude squared over $z$ we get the number $n$ of photons in the soliton as

$$\int dz \, |a_c(z,t)|^2 = \int A^2 \sec h^2\left(\frac{z}{\xi}\right) dz = 2A^2\xi = n \quad . \tag{28}$$

We notice according to Eq. (28) that the number of photons in the soliton is proportional to $A^2$ representing the normalization constant squared. The parameter $\xi$ is related to the pulse shape. For larger values of $\xi$ the soliton pulse becomes narrower with a larger amplitude at the pulse center and larger number of photons. The one-soliton solution of the NLS equation represents a balance between linear dispersion, which tends to break up the soliton wave packet, and a focusing effect of the cubic nonlinearity, produced by the interaction of the wave with itself.

Returning back to the QM equation (24) for $\hat{a}(z,t)$ its solution under the above assumptions can be represented as

$$\hat{a}(z,t) = \hat{a} \, g(z,t) \quad where \quad g(z,t) = \exp\left[i\left(\frac{KA^2}{2}t\right)\right] \sec h\left(\frac{z}{\xi}\right) \tag{29}$$

We find the interesting property that the time dependence of the one soliton solution, for both the classical solution (26) and the QM solution (29), is separated from its space $z$ dependence. These solutions can be illuminated by changing their space $z$ dependence into the wavevector $k$ dependence using the Fourier transform

$$F(k) = \frac{1}{\sqrt{2\pi}} \int_{-\infty}^{\infty} \exp(-ikz) \operatorname{sech}\left(\frac{z}{\xi}\right) dz = \xi \sqrt{\frac{\pi}{2}} \operatorname{sech}\left(\frac{\pi k \xi}{2}\right) \quad , \tag{30}$$

Then the QM solution of Eq. (29) is transformed into

$$\hat{a}(z,t) = \hat{a}\, g(k,t) \quad \text{where} \quad g(k,t) = \exp\left[i\left(\frac{KA^2}{2}t\right)\right] F(k) \quad . \tag{31}$$

Eq. (31) has a quite simple explanation showing that in the one-soliton pulse we have a certain distribution of wavevectors producing the wavepacket described in Eq. (29). All these wavevectors have the same time dependence represented by the term $\exp\left[i\left(\frac{KA^2}{2}t\right)\right]$. One needs, however, to take into account that the rapid frequency dependence $\exp(-i\omega_0 t)$ has been eliminated by using Eq. (19) (omitting for simplicity of notation the tilde on operator $\hat{a}$). Also the relative simple forms of Eqs. (26-31) are obtained under the above simplifying assumptions.

Both the classical Eq. (26) and the QM Eq. (29) give information for the one-soliton phases and space distributions. We can use a certain integration over the $z$ dependence of the soliton exchanging the classical solution of Eq. (26) into

$$a_c(t) = \sqrt{2\xi}\, A \exp\left[i\frac{KA^2}{2}t\right] \quad , \tag{32}$$

so that in agreement with Eq. (28) we will get

$$|a_c(t)|^2 = n \quad . \tag{33}$$

The nonlinear Hamiltonian of Eq. (23) has led to the phase term in Eq. (32).

The QM operator corresponding to the complex classical amplitude of Eq. (32) is given by

$$\hat{a} = \hat{a}(0)\sqrt{2\xi}\, A \exp\left(i\frac{KA^2}{2}t\right) \qquad (34)$$

A simple QM description of the photon number distribution in the one-soliton pulse can be obtained by assuming that the quantum analog of Eq. (32) is given by the coherent state

$$|\alpha(t)\rangle = \left|\alpha(0)\exp\left[i\left(\frac{KA^2}{2}t\right)\right]\right\rangle \quad where \quad \alpha(0) = \sqrt{2\xi}\, A \quad . \qquad (35)$$

The photon number distribution of the one-soliton state becomes then [22]:

$$|\alpha(t)\rangle = \frac{\left\{\sqrt{2\xi}\, A\exp\left[ikA^2 t/2\right]\right\}^n}{\sqrt{n!}} \exp\left(-\xi A^2\right)|n\rangle \qquad (36)$$

For each number state $|n\rangle$ in the coherent photon distribution an additional phase given by $\exp\left[ikA^2 tn/2\right]$ is introduced. The phase $KA^2 tn/2$ is proportional to the photon number and to $A^2$ (see Eq.(28)) and increases linearly with time. Similar phases are known to be obtained by the ordinary Kerr interactions[18-20].

     One should take care about the fact that the analysis of photon statistics for solutions of NLS equation by the use of quantum field theories becomes extremely complicated. We have simplified such analysis by using certain analogies between the classical and QM solutions and have obtained two important QM new results: 1) The photon number distribution in the one-soliton state is approximately that of a coherent state. (Also in [8,21] *the photon number uncertainty* has been found to be that of a Poisson distribution). By repeating many times the preparation of the one-soliton under the same conditions and measuring in each time the total photon number such distribution might be observed. 2) The present analysis shows that there are Kerr phases which are proportional to the one-soliton photon numbers and to the normalization constant $A^2$. Such phases might be observed by interference experiments [18-20,22].

## 4. Integrability condition for nonlinear equations applied to NLS equation

There is a lot of literature in which soliton solutions of nonlinear equations have been developed using the inverse scattering methods [1]. In particular Zakharov and Shabat [2] have developed such methods for treating the NLS equation. Our interest in the present Section is, however, only to show that the integrability of NLS equation can be related to the *compatibility-condition* between Hamiltonian and Momentum matrices. We limit the discussion to the integrability problem and we do not discuss scattering theories which are mainly *classical* as our interest in the present analysis is to analyze *quantum* effects which will be related to Quantum Optics methods.

Suppose that $u(z,t)$ satisfies some nonlinear evolution equation of the form

$$\frac{\partial u(z,t)}{\partial t} = N(u(z,t)) \tag{37}$$

where $N(u(z,t))$ is a nonlinear operator which is independent of $t$ but involves $z$ and derivatives relative to $z$. In particular we relate the analysis in the present Section to solution for the one-soliton state of NLS equation which has been presented in many works. For example such solution has been given by Zakharov and Shabat ([2] Eq. (6)) as:

$$u(z,t) = A_0 \frac{\exp\{-4i(\xi^2 - \eta^2)t - 2i\xi x + i\varphi\}}{\cosh[2\eta(x - x_0) + 8\eta\xi t]}, \tag{38}$$

where $A_0$ is a normalization constant. This solution is characterized by 4 constants $\eta, \xi, x_0$ and $\varphi$. By assuming in Eq. (38) $\xi = x_0 = \varphi = 0$ and $2\eta \equiv \frac{1}{\xi}$ this equation has been reduced in the previous Section to Eq.(26) with the constraint of Eq. (27) (assuming $C = 2$). Of course, our previous classical solution can be generalized if we relax the above special conditions.

Let us assume that we have a two dimensional wavefunction depending both on $z$ and $t$

$$\vec{\psi}(z,t) = \begin{pmatrix} \psi_1(z,t) \\ \psi_2(z,t) \end{pmatrix} \qquad . \tag{39}$$

Suppose that $|\vec{\psi}(z,t)\rangle$ satisfy the equation :

$$\frac{\partial}{\partial t}\begin{pmatrix} \psi_1(z,t) \\ \psi_2(z,t) \end{pmatrix} = \mathrm{H}\begin{pmatrix} \psi_1(z,t) \\ \psi_2(z,t) \end{pmatrix} \qquad , \tag{40}$$

where H is a two dimensional matrix which can include a function of $u(z,t)$ and its derivatives. Suppose also that $|\vec{\psi}(z,t)\rangle$ satify the equation

$$\frac{\partial}{\partial z}\begin{pmatrix} \psi_1(z,t) \\ \psi_2(z,t) \end{pmatrix} = \mathrm{M}\begin{pmatrix} \psi_1(z,t) \\ \psi_2(z,t) \end{pmatrix} \qquad , \tag{41}$$

where M is a two dimensional matrix which also can include a function of $u(z,t)$ and its derivatives. We assume the *Compatibility-Condition*

$$\frac{\partial}{\partial z}\left(\frac{\partial}{\partial t}\vec{\psi}(z,t)\right) = \frac{\partial}{\partial t}\left(\frac{\partial}{\partial z}\vec{\psi}(z,t)\right) \qquad , \tag{42}$$

where on the left side of this equation we perform first the derivative of $\vec{\psi}(z,t)$ first according to $t$ and afterwards according to $z$ while on the right side we inverse the order of these derivatives. Using Eqs, (40-41) we get

$$\frac{\partial}{\partial z}\left(\frac{\partial}{\partial t}\vec{\psi}(z,t)\right) = \frac{\partial}{\partial z}\left(\mathrm{H}\vec{\psi}(z,t)\right) = \left(\frac{\partial}{\partial z}\mathrm{H} + \mathrm{HM}\right)\vec{\psi}(z,t) \qquad , \tag{43}$$

$$\frac{\partial}{\partial t}\left(\frac{\partial}{\partial z}\vec{\psi}(z,t)\right) = \frac{\partial}{\partial t}\left(\mathrm{M}\vec{\psi}(z,t)\right) = \left(\frac{\partial}{\partial t}M + \mathrm{MH}\right)\vec{\psi}(z,t) \qquad . \tag{44}$$

Substituting Eqs. (43-44) into Eq. (42) we get the *Compatibility Condition*

$$\mathrm{HM} - \mathrm{MH} - \frac{\partial}{\partial t}\mathrm{M} + \frac{\partial}{\partial z}\mathrm{H} = 0 \tag{45}$$

The matrix operators M and H are known as *Lax pairs* [24]. The idea is that by using special forms for the matrices M and H which satisfy the compatibility equation (45) they lead to *integrable* nonlinear equation. We demonstrate here such an approach for the NLS equation. For this purpose we define, for example:

$$M = \begin{pmatrix} -i\varsigma & u(z,t) \\ -u^*(z,t) & i\varsigma \end{pmatrix} \quad , \tag{46}$$

$$H = \begin{pmatrix} -i|u(z,t)|^2 + 2i\varsigma^2 & -i\dfrac{du(z,t)}{dz} - 2u(z,t)\varsigma \\ -i\dfrac{du^*(z,t)}{dz} + 2u^*(z,t)\varsigma & i|u(z,t)|^2 - 2i\varsigma^2 \end{pmatrix} \tag{47}$$

where $\varsigma$ is a real parameter independent of time $t$ and space z.

Substituting Eqs (46) and (47) in the *compatibility equation* (45) we get after straightforward calculations

$$\begin{pmatrix} 0 & \dfrac{\partial u}{\partial t} + i\dfrac{\partial^2 u}{\partial z^2} + 2i|u|^2 u \\ \dfrac{\partial u^*}{\partial t} - i\dfrac{\partial^2 u}{\partial z^2} - 2i|u|^2 u^* & 0 \end{pmatrix} = \begin{pmatrix} 0 & 0 \\ 0 & 0 \end{pmatrix} \quad . \tag{48}$$

While the diagonal elements in the *compatibility equation* vanish in a trivial way the vanishing of the nondiagonal elements lead to NLS equation (up to some normalization constants):

$$\frac{\partial u(z,t)}{\partial t} + i\frac{\partial^2 u(z,t)}{\partial z^2} + 2i|u(z,t)|^2 u(z,t) = 0 \quad , \tag{49}$$

and to the complex conjugate of this equation. It is quite interesting to note that while Eqs. (40) and (41) depend on the parameter $\varsigma$ included in Eqs. (46) and (47), the *compatibility-condition* leads to the NLS equations in Eq. (48) which are independent of this parameter.

Eqs. (40) and (41) have a simple geometric interpretation as these equations describe *connections* on a two-dimensional *vector bundle* over the $(z,t)$ plane [25]. Eq. (41) describes how to 'parallel translate' the vector $\vec{\psi}(z,t)$ in the $z$-direction and Eq. (40)

describes how to 'parallel translate' $\vec{\psi}(z,t)$ in the $t$-direction. The matrices M and H are the 'connection coefficients'. A connection is defined to have a *zero curvature* if parallel translation between two nearby points is independent of the path connecting the two points. Therefore the *compatibility-condition* represents the *zero curvature* condition for the *integrability* of the nonlinear equation. We have shown that there is a two dimensional wavefunction $\vec{\psi}(z,t)$ satisfying both Eqs. (40) and (41) with the *compatibility-condition* which is reduced to the NLS equation.

One should notice that Eq. (40) can be changed into

$$i\frac{\partial}{\partial t}\begin{pmatrix}\psi_1(z,t)\\ \psi_2(z,t)\end{pmatrix} = i\mathrm{H}\begin{pmatrix}\psi_1(z,t)\\ \psi_2(z,t)\end{pmatrix} \quad , \tag{50}$$

where $i\mathrm{H}$ becomes Hermitian so that Eq. (50) gets the form of Schrodinger equation (with $\hbar = 1$). In the same way Eq. (41) can be changed into the form

$$-i\frac{\partial}{\partial z}\begin{pmatrix}\psi_1(z,t)\\ \psi_2(z,t)\end{pmatrix} = -i M\begin{pmatrix}\psi_1(z,t)\\ \psi_2(z,t)\end{pmatrix} \quad , \tag{51}$$

where $-iM$ becomes Hermitian so that Eq. (51) gets the form of the *propagation momentum* matrix [12-17] (with $\hbar = 1$).

Eq. (51) leads to Hilbert space with a quantum basis of two orthonormal wavefunctions with inner products. Such basis of wavefunctions can be used for giving the initial state where its time dependence will be given by Eq. (50). However, in Eqs. (46) and (47) we have assumed that *we have already the solution* of $u(z,t)$. The *inverse scattering theories* go in inverse direction where by using quite complicated methods one finds the soliton solution from *scattering matrices*. We have given the analysis in the present Section only for clarifying the integrability of the *nonlinear equation* which is obtained by a soliton.

We do not develop scattering equations (see e.g. [1-2, 6-7]) and their relations to solitons as such theories are usually classical and our aim in the present study is to treat QM effects. For analyzing Kerr phases effects it will be more efficient to remain with the use of the *coupled Hamiltonian-Momentum operator* as has been developed in Section 3. The discussion given in the present Section, however, illuminates some important properties about the integrability of the NLS equation.

## 5. Summary and conclusions

In the present study we have analyzed classical and QM effects for electromagnetic (EM) waves which satisfy the one-soliton non-linear Schrodinger (NLS) equation in a dispersive wave guide using quantum optics methods. We have treated one mode of the EM field which includes a coupling between momentum and frequencies due to dispersion relation. For this purpose *a coupled Hamiltonian-Momentum operator* with equal-space commutation relations (CR) has been used. While most theories about solitons are purely classical [1-2,6-7,10-11] the present treatment of the one-soliton state includes QM effects which are developed in analogy with the classical theory. It has been shown that by the Kerr interaction in the soliton, a coherent photon number distribution is obtained with corresponding Kerr quantum phases. The present analysis is different from that of quantum field theories [4-5,9] and from that of other QM theories on solitons [8,21] as it is more directly related to quantum optics theories of Kerr interactions in other systems [16-20].

We have analyzed the relation between the integrability of nonlinear equations and the *compatibility-condition* for two equations where one of these equations is related to Hamiltonian and the other to Momentum matrices. It has been shown that by choosing special equations the *compatibility-condition* leads to NLS equation and we have discussed the integrability condition for this equation.